\documentclass[aps, pre,amsmath,amssymb,twocolumn,floatfix]{revtex4}
\usepackage[usenames,dvipsnames]{color}
\usepackage[pdftex]{graphicx}
\usepackage{epsfig}
\input epsf

\newcommand {\dx}{{\mathrm d}\mathbf{x}}

\newcommand {\xx}{\mathbf{x}}
\newcommand{\vx}{{\bf x}}\newcommand{\evalAt}[1]{\bigg |_{#1}}

\newcommand {\kk}{\mathbf{k}}

\newcommand{\paDir}[2]{\frac{\partial #1}{\partial #2}}
\newcommand{\fnDir}[2]{\frac{\delta #1}{\delta #2}}

\graphicspath{{./Fig/}} 

\newcounter{countcoaut}

\begin{document}

\title{\bf { Solidification} fronts in supercooled liquids: {how rapid fronts can lead to disordered glassy solids}}

\author{A.J.~Archer}
\author{M.J.~Robbins}
\author{U.~Thiele}
\affiliation{Department of Mathematical Sciences, Loughborough University, Loughborough LE11 3TU, United Kingdom}
\author{E.~Knobloch}
\affiliation{Department of Physics, University of California at Berkeley, Berkeley, CA 94720}

\begin{abstract} 

We determine the speed of a crystallisation (or more generally, a solidification) front as it advances into the uniform liquid phase after the system has been quenched into the crystalline region of the phase diagram. We calculate the front speed by assuming a dynamical density functional theory model for the system and applying a marginal stability criterion. Our results also apply to phase field crystal (PFC) models of solidification. As the solidification front advances into the unstable liquid phase, the density profile behind the advancing front develops density modulations and the wavelength of these modulations is a dynamically chosen quantity. For shallow quenches, the selected wavelength is precisely that of the crystalline phase and so well-ordered crystalline states are formed. However, when the system is deeply quenched, we find that this wavelength can be quite different from that of the crystal, so that the solidification front naturally generates disorder in the system. Significant rearrangement and ageing must subsequently occur for the system to form the regular well-ordered crystal that corresponds to the free energy minimum. Additional disorder is introduced whenever a front develops from random initial conditions. We illustrate these findings with results obtained from the PFC.

\end{abstract} 
\maketitle 
\epsfclipon  


\section{Introduction}

It is important to understand the formation kinetics of a solid from the liquid phase when it is cooled below its freezing temperature $T_f$, because the microscopic structure of the solid can depend strongly on the formation pathway. If the liquid is only slightly cooled below $T_f$, then the solid forms via nucleation and growth \cite{Oxtoby92}, generally leading to well-ordered crystalline solids. Depending on the material and the degree of cooling below $T_f$, the formation of dendrites and other complex microstructures is possible \cite{BWBK02, TTG11, TGTDP11}. When the liquid is rapidly quenched (supercooled) to a temperature sufficiently far below $T_f$ there is no nucleation barrier against the liquid forming a solid. In this situation, the solid that is formed can be { amorphous, with little or no long-range order}, rather than a regular ordered crystal.

Classical density functional theory (DFT) \cite{Evan79,Evan92} is a widely used microscopic theory capable of describing equilibrium aspects of melting, freezing and also the interfaces between the liquid and solid phases \cite{singh91, lowen94}. In conjunction with the recent development of a dynamical density functional theory (DDFT) \cite{MT99,MT00,archer04,archer04b}, which is a theory that requires as input the free energy functionals from equilibrium DFT, these theories have been shown to be able to describe the dynamics of crystal formation \cite{TBVL09}. A related approach, that has been developed and studied extensively over the last decade or so, is the phase field crystal (PFC) \cite{EG04, BGE06, EPBSG07, BRV07, BEG08a, BEG08, MKP08, TBVL09, GE11, TTG11, TGTDP11,RATK12} approach for modeling the atomic structure of crystalline materials. The PFC may be derived from the DDFT by assuming a (local) gradient expansion approximation for the Helmholtz free energy functional for the system and linearizing the density-dependent mobility pre-factor in the DDFT equation \cite{EPBSG07, TBVL09}. DFT and DDFT are theories for the one body density distribution $\rho(\xx)$ of the particles in the system. These theories essentially treat the solid phase as an inhomogeneous liquid, in which the density profile consists of an array of density peaks, each corresponding to a localized particle, in contrast to the liquid phase which has a uniform density distribution. The PFC is a theory for an order parameter profile $\phi(\xx)$, which in a similar manner takes a constant value in the liquid phase and forms an array of peaks in the solid phase.

In this paper we consider a simple liquid that has been rapidly quenched to a temperature well below $T_f$ and develop a theory for how the solidification front propagates into the unstable liquid. We base our analysis on the DDFT and PFC models. DDFT predicts that the time evolution of the one-body density $\rho(\xx,t)$ of a system of particles is governed by the following equation \cite{MT99,MT00,archer04,archer04b}:
\begin{equation}
\frac{\partial \rho(\xx,t)}{\partial t}=\Gamma \nabla \cdot \left[ \rho(\xx,t)\nabla \frac{\delta F[\rho]}{\delta \rho(\xx,t)} \right],
\label{eq:DDFT}
\end{equation}
where $\Gamma$ is a (constant) mobility coefficient and $F[\rho(\xx,t)]$ is the equilibrium fluid Helmholtz free energy functional:
\begin{eqnarray} \notag
F[\rho(\xx,t)] \,=\, \beta^{-1} \int \dx \,\rho(\xx,t)[\ln(\rho(\xx,t)\Lambda^3)-1] \\
\quad +\, F_\mathrm{ex}[\rho(\xx,t)] \,+ \, \int \dx \,V_\mathrm{ext}(\xx,t) \rho(\xx,t).
\label{eq:freeenergy}
\end{eqnarray}
The first term is the ideal gas free energy; $\Lambda$ is the thermal de Broglie wavelength and $\beta=1/k_BT$ is the inverse temperature. The second term $F_\mathrm{ex}$ is the excess contribution and the final term is the contribution from the external potential $V_\mathrm{ext}(\xx,t)$. The DDFT may be derived from the Smoluchowski (Fokker-Planck) equation for a system of interacting Brownian particles with overdamped stochastic equations of motion, by assuming that the two-body correlations in the non-equilibrium fluid are the same as those in an equilibrium fluid with the same one-body density profile \cite{archer04}. Moreover, for dense atomic or molecular fluids, in which the equations of motion for the particles are, of course, Newton's equations of motion, one can argue \cite{archer17, archer25} that Eqs.\ \eqref{eq:DDFT} and \eqref{eq:freeenergy} still provide a reasonable approximation for the dynamics of the system, particularly when it is not too far from equilibrium.

This paper is laid out as follows: In Sec.\ \ref{sec:disp_rel} we consider the stability of a uniform liquid with number density $\rho(\xx)=\rho_0$ and obtain the dispersion relation for the growth/decay of small amplitude harmonic density perturbations. We then approximate this relation and obtain a simple expression which coincides with the dispersion relation that one obtains from considering the PFC theory. { In Sec.\ \ref{sec:front_speed} we employ the marginal stability hypothesis to compute from this dispersion relation the speed of a solidification front advancing into a linearly unstable uniform liquid \cite{GE11}.} We also make an expansion in a certain small parameter related to undercooling, in order to obtain an analytical expression for the speed $c$ of the solidification front. We find that the wavelength $\lambda$ of the density modulations which develop in the system as the solidification front advances, is not necessarily equal to the lattice spacing $\lambda_c$ of the equilibrium crystal that the system seeks to form. This is because the length $\lambda$ is a dynamically selected (non-equilibrium) quantity. When the liquid is only weakly supercooled { into the linearly unstable region}, then $\lambda \approx \lambda_c$ and one should expect a regular crystal to form easily. However, when the system is deeply supercooled, then $\lambda\neq\lambda_c$ and one should expect the formation of a regular crystal to be frustrated and the structure that is initially formed behind the advancing solidification front to be somewhat disordered (amorphous). In Sec.\ \ref{sec:PFC_res} we confirm this conclusion, i.e., that a deep quench leads initially to the formation of solids with greater disorder, using numerical simulations of the PFC model in two spatial dimensions. We also show that the transverse filamentation of the stripe pattern nucleated by the advancing front is a consequence of the random initial conditions we employ. { In Sec.\ \ref{sec:conc}, we draw our conclusions and discuss the applicability of our PFC results to understanding real materials.}

\section{Dispersion relation}
\label{sec:disp_rel}

\begin{figure}
\includegraphics[width=0.95\columnwidth]{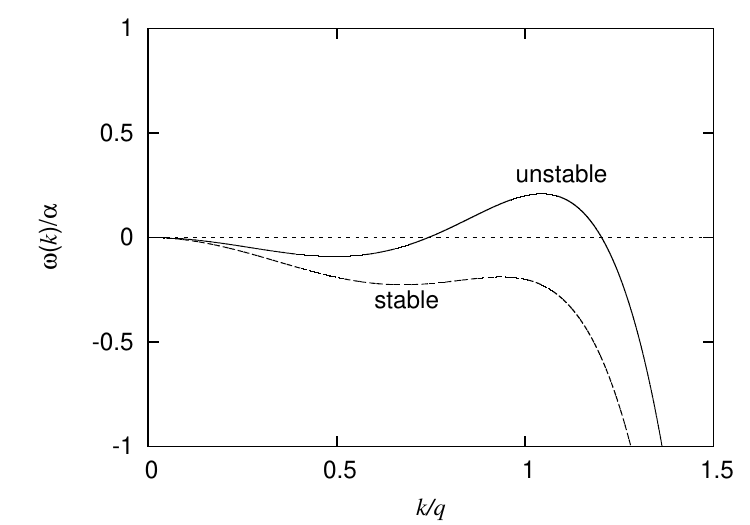}
\caption{\label{fig:disp_rel}
Sketch of the approximate dispersion relation $\omega(k)$ in Eq.\ \eqref{eq:disp_rel_PFC}. When the dispersion relation takes the form labelled `stable' the uniform liquid is linearly stable ($\Delta>0$, dashed line). In the case labelled `unstable' the uniform fluid is linearly unstable and density modulations with wave number $k\approx q$ grow in amplitude, leading to the formation of the solid phase ($\Delta<0$, solid line).}
\end{figure}

We consider a bulk fluid, where the external potential $V_\mathrm{ext}(\xx,t)=0$ in Eq.\ \eqref{eq:freeenergy} and consider small density fluctuations $\tilde{\rho}(\xx,t)=\rho(\xx,t)-\rho_0$ about the bulk fluid density $\rho_0$. We have in mind that we are considering a homogeneous fluid which has been rapidly quenched to the region of the phase diagram where the crystal is the equilibrium phase. In the following derivation of the dispersion relation for the growth/decay of harmonic density fluctuations $\tilde{\rho}(\xx,t)$, we initially follow Ref.\ \cite{archer04}. From Eqs.\ \eqref{eq:DDFT} and \eqref{eq:freeenergy} we obtain:
\begin{eqnarray}\notag
\frac{1}{D} \frac{\partial \tilde{\rho}(\xx,t)}{\partial t}
\, =\, \nabla^2 \tilde{\rho}(\xx,t)
\,-\, \rho_0 \nabla^2 c^{(1)}(\xx,t) \\
-\, \nabla . [ \, \tilde{\rho}(\xx,t) \nabla c^{(1)}(\xx,t) \, ],
\label{eq:Smoluchowski_2}
\end{eqnarray}
where the diffusion coefficient $D=\Gamma/\beta$ and $c^{(1)}(\xx,t)=-\beta\delta F_\mathrm{ex}/\delta \rho$ is the one-body direct correlation function \cite{Evan79,Evan92}. We linearise Eq.\ (\ref{eq:Smoluchowski_2}) in $\tilde{\rho}$ by Taylor expanding $c^{(1)}$ about the bulk fluid value, giving
\begin{equation}
c^{(1)}(\xx) \, =\, c^{(1)}(\infty)
\,+\, \int \dx' \,\frac{\delta c^{(1)}(\xx)}{\delta
\rho(\xx')} \Bigg \vert_{\rho_0} \tilde{\rho}(\xx',t)
\, +\, {\cal O} (\tilde{\rho}^2),
\label{eq:c_1_expansion}
\end{equation}
where $c^{(1)}(\infty) \equiv c^{(1)}[\rho_0]=-\beta \mu_\mathrm{ex}$ and $\mu_\mathrm{ex}$ is the excess chemical potential. Note also that
\begin{eqnarray}\notag
\frac{\delta c^{(1)}(\xx)}{\delta \rho(\xx')}
&=&-\beta \frac{\delta^2 F_\mathrm{ex}[\rho]}{\delta \rho(\xx') \delta \rho(\xx)}\\
&=& c^{(2)}(\xx,\xx')\notag \\
&=& c^{(2)}(|\xx-\xx'|;\rho_0) 
\label{eq:c_2}
\end{eqnarray}
for a homogeneous fluid of spherically symmetric particles. For an equilibrium system $c^{(2)}(|\xx-\xx'|;\rho_0)$ is the Ornstein-Zernike direct pair correlation function of the fluid with density $\rho_0$. Substituting Eq.\ \eqref{eq:c_1_expansion} into Eq.\ \eqref{eq:Smoluchowski_2}, we obtain \cite{archer04}:
\begin{eqnarray}
\notag
\frac{1}{D} \frac{\partial \tilde{\rho}(\xx,t)}{\partial t}
\, =\, \nabla^2 \tilde{\rho}(\xx,t)\,
\hspace{4cm} \\
-\, \rho_0 \nabla^2 \left[ \, \int \, \dx' c^{(2)}(|\xx-\xx'|;\rho_0)
\tilde{\rho}(\xx',t) \, \right]+{\cal O}(\tilde{\rho}^2) .
\label{eq:Spin_dec_eq}
\end{eqnarray}
We now assume that the density fluctuation is of the form $\tilde{\rho}(\xx,t)=\epsilon\exp(\omega t+i\kk.\xx)$, where $\epsilon$ is a small amplitude and the dispersion relation $\omega(k)$, where $k=|\kk|$, is yet to be determined. From Eq.\ \eqref{eq:Spin_dec_eq} we obtain:
\begin{equation}
\frac{\omega}{D}\tilde{\rho}(\xx,t)
\, =\, -k^2 \tilde{\rho}(\xx,t) \,+\, \rho_0 k^2 \, \hat{c}(k) \tilde{\rho}(\xx,t)+{\cal O}(\tilde{\rho}^2),
\label{eq:Spin_dec_eq_2}
\end{equation}
where $\hat{c}(k)=\int \dx \exp(-i \kk.\xx) c^{(2)}(x;\rho_0)$ is the Fourier transform of the pair direct correlation function. Note that for an {\em equilibrium} fluid, at a state point outside the spinodal, $S(k) \equiv (1 \,- \, \rho_0 \hat{c}(k))^{-1}$ is the static structure factor. Linearising Eq.\ \eqref{eq:Spin_dec_eq_2} we obtain the dispersion relation:
\begin{equation}
\omega(k) = - D k^2 [1-\rho_0\hat{c}(k)].
\label{eq:disp_rel}
\end{equation}
It is clear that small density fluctuations only grow in amplitude if for some wave numbers $k$ we have $\omega(k)>0$. Crystallisation occurs when the system is unstable against periodic density modulations, which occurs when $\omega(k)>0$ for a band of wave numbers about $k\approx q$, where $q\neq 0$. The dispersion relation $\omega(k)$ for an unstable system is of the form sketched using the solid line in Fig.\ \ref{fig:disp_rel}. Note that crystals may form before the system becomes linearly unstable; however, in this case the crystal must be nucleated. { Furthermore, if the fluid state falls within the solid-liquid coexistence region, then} the crystal front will not advance indefinitely: it will grow until it has removed sufficient material from the surrounding fluid to produce phase coexistence between the liquid and the crystal. In this case, the crystal forms a `localised state'. PFC results for this situation may be seen in Refs.\ \cite{TTG11,TGTDP11,RATK12}. 

In the following we assume that the speed with which the crystallisation front advances into the unstable liquid corresponds to the marginal stability criterion \cite{DL83,benjacob85,HuMo90}. Specifically, we suppose that the unstable liquid state is characterized by a dispersion relation $\omega=\omega(k)$. In the frame in which the front is stationary, the dispersion relation becomes $\omega=ick+\omega(k)\equiv\Omega(k)$, where $c$ is the speed of the front. In this frame the following relations hold
\begin{eqnarray}\label{eq:speed1}
\frac{d\Omega}{dk}=0\label{eq:speed2a}\\
\Re(\Omega)=0,
\label{eq:speed2}
\end{eqnarray}
corresponding to the presence of a double root of $\omega=\Omega(k)$ in the complex $k$ plane together with the requirement that the perturbation neither grows nor decays. Since $\Im(\Omega)\ne 0$ in general, the wavetrain left behind by the moving front has a well-defined frequency in the frame of the front.

The above conditions are equivalent to three conditions which are to be solved for the speed $c$ of the crystallisation front together with the associated complex wave number $k\equiv k_r+ik_{i}$. The resulting density profile at or before the front has the form $\tilde{\rho}(\xx,t) = \tilde{\rho}_\mathrm{front}(\xi,t)$, where $\xi\equiv x-ct$ represents the position relative to the moving front and $\tilde{\rho}_\mathrm{front}(\xi,t) \sim \exp(-k_{i}\xi)\sin(k_r\xi+\Im(\Omega)t)$. Thus $k_r$ is the wave number of the growing perturbation, i.e., the wave number before the front, while $k_{i}$ represents the spatial decay (growth) of the perturbation in the forward (backward) direction. In contrast, the pattern left behind by the front is a fully nonlinear periodic state with wave number $k^*$, say. In the absence of phase slips such a state takes the form $\rho(k^*\xi+\Im(\Omega)t)$, i.e., a wave that travels backwards relative to the front with frequency $\Im(\Omega)$; with no phase slips this frequency is {\it identical} to the frequency ahead of the front and so can be computed from the marginal stability calculation. In view of the gradient structure of Eq.~(\ref{eq:DDFT}) this solution must be stationary in the laboratory frame. Thus $\rho(k^*x-k^*ct+\Im(\Omega)t)$ must be independent of the time $t$, implying that \cite{benjacob85}
\begin{equation} 
k^*=\frac{1}{c}\Im(\Omega)=k_r+\frac{1}{c}\Im[\omega(k)].\label{eq:kbehind}
\end{equation} 
This equation expresses the conservation of nodes. Note that $k^*$ differs in general from the marginal stability wave number $k_r$.

To obtain the crystallisation front speed $c$, one assumes an approximation for $F_\mathrm{ex}$ in Eq.\ \eqref{eq:freeenergy} to obtain an expression for $\hat{c}(k)$ \footnote{Ref.\ \cite{archer04}, for example, gives an approximation for this quantity for a fluid particles interacting via a pair potential that has a hard-sphere plus attractive Yukawa tail.}, and hence the approximate dispersion relation $\omega(k)$. With this input Eqs.~\eqref{eq:speed2a} and \eqref{eq:speed2} may be solved (numerically) for $c$, $k_r$ and $k_{i}$ and the wave number of the deposited pattern evaluated using \eqref{eq:kbehind}. Under certain conditions an approximate solution to this problem may be obtained analytically, as shown next.

\section{Solidification front speed}
\label{sec:front_speed}

\subsection{Approximate dispersion relation}

To compute the front speed we first derive an approximation to the dispersion relation by expanding $\hat{c}(k)$ in powers of $k$. In order to capture the peak at $k\approx q$ in the dispersion relation, one must retain at least terms up to $O(k^4)$ in $\hat{c}(k)$. Thus we write
\begin{equation}
\hat{c}(k) \approx c_0+c_2k^2+c_4k^4
\label{eq:c_approx}
\end{equation}
and suppose that $c_4<0$. This approximation corresponds to making a gradient expansion of the free energy $F_\mathrm{ex}[\rho]$ and retaining only terms up to and including the terms $\sim -[\nabla^2\rho(\xx,t)]^2$ \footnote{The coefficients $c_i$ in Eq.\ \eqref{eq:c_approx} are in general functions of the fluid density and are related to the coefficients in a gradient expansion of the free energy. For example, $c_2(\rho_0)=-\frac{1}{6}\int \dx \,x^2 c^{(2)}(x;\rho_0)=-2\beta f_2(\rho_0)$, where $f_2$ is the coefficient of the gradient squared term. For further details see Ref.\ \cite{Evan79} and Appendix A.}. Substituting Eq.\ \eqref{eq:c_approx} into Eq.\ \eqref{eq:disp_rel}, we obtain
\begin{equation}
\omega(k) = -\alpha k^2[\Delta+(q^2-k^2)^2],
\label{eq:disp_rel_PFC}
\end{equation}
where $\alpha=-\rho_0c_4D$, $q^2=-c_2/2c_4$ and $\Delta=(\rho_0c_0-1)/\rho_0c_4-(c_2/2c_4)^2$. The uniform fluid thus becomes linearly unstable for $\Delta<0$; i.e., the stable dispersion curve in Fig.\ \ref{fig:disp_rel} corresponds to a case when $\Delta>0$ and the unstable curve is for $\Delta<0$. Thus the magnitude of the parameter $\Delta$ indicates how deep one has quenched into the region of the phase diagram where the uniform liquid is linearly unstable.

Note that the dispersion relation in Eq.\ \eqref{eq:disp_rel_PFC} is exactly that which one obtains when considering the PFC model for the order parameter $\phi(\xx,t)=[\rho(\xx,t)-\rho_0]/\rho_1$, where $\rho_1$ is a constant. The PFC model may be derived from the DDFT by assuming a gradient expansion in $F_\mathrm{ex}$ and expanding the free energy in powers of $\phi$ and then linearising certain terms \cite{EPBSG07,TBVL09, Robb12}, obtaining
 \begin{equation}
\frac{\partial \phi(\xx,t)}{\partial t}=\alpha \nabla^2 \frac{\delta F[\phi]}{\delta \phi(\xx,t)} .
\label{eq:DDFT_PFC}
\end{equation}
Here $\alpha$ is a mobility coefficient and the free energy functional
\begin{eqnarray}
F[\phi]\equiv \int \dx \left[\frac{\phi}{2}[r+(q^2+\nabla^2)^2]\phi+\frac{\phi^4}{4}\right].
\label{eq:freeenergy_PFC}
\end{eqnarray}
Details of this derivation are contained in Appendix A. For the PFC model, we find that the uniform state $\phi(\xx,t)=\phi_0$ (corresponding to the liquid) is linearly unstable when the undercooling parameter $r<-3\phi_0^2$. Thus, in this model we have $\Delta=r+3\phi_0^2$ and $\Delta<0$ represents the undercooled liquid state. 

For the one-dimensional PFC model the marginal stability analysis described above was performed in Ref.~\cite{GE11}. In the remainder of this paper we extend the predictions of this approach both analytically and numerically, and compare them with results from numerical simulations in one and two dimensions.

\subsection{The front speed}

\begin{figure*}

\includegraphics[width=3in]{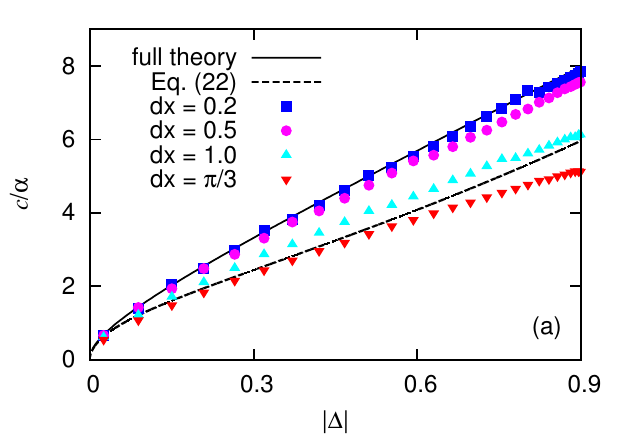}
\includegraphics[width=3in]{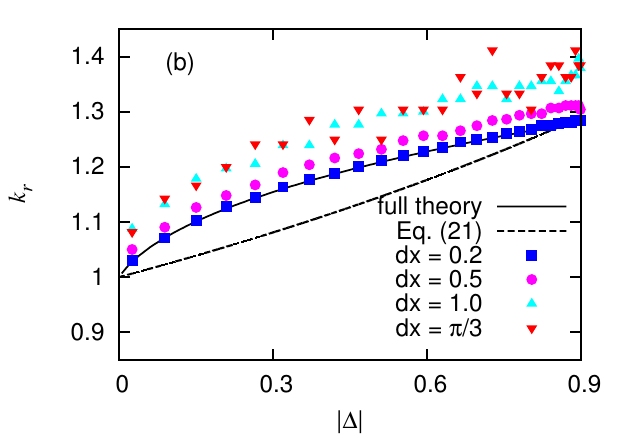}\\
\includegraphics[width=3in]{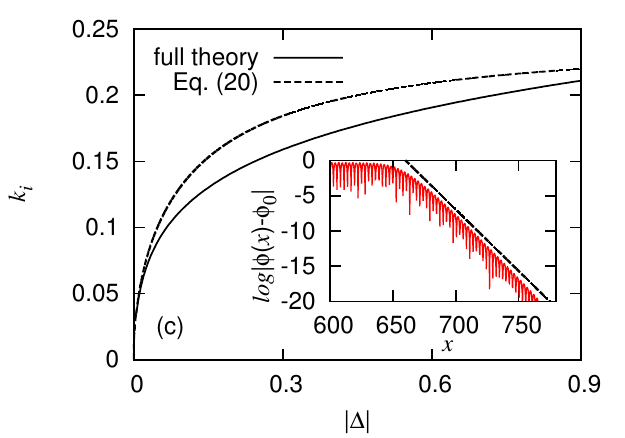}
\includegraphics[width=3in]{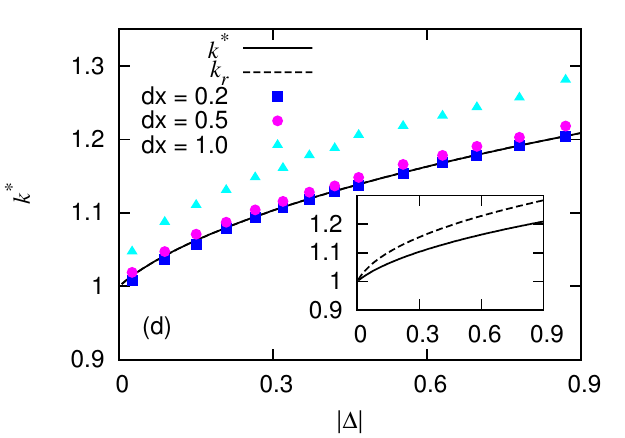}
\caption{\label{fig:speed}
(Colour online) (a) The crystallization front speed $c$ as a function of $|\Delta|$, for $q=1$, $\alpha=1$. The solid line is the result from solving the full theory [Eqs.\ \eqref{eq:speed1}, \eqref{eq:speed2} and \eqref{eq:disp_rel_PFC}], and the dashed line is the analytical approximation in Eq.\ \eqref{eq:analytic_speed}. Note that in the PFC model $\Delta=r+3\phi_0^2$. The symbols correspond to numerical results obtained for the front speed when the PFC Eqs.\ \eqref{eq:DDFT_PFC} and \eqref{eq:freeenergy_PFC} are discretized with various values for the spatial grid spacing $dx$, as indicated in the figure. In (b) we display the corresponding results for $k_r$ and in (c) for $k_{i}$. In the inset of (c) we display the order parameter profile plotted as $\log |\phi(x)-\phi_0|$ versus $x$, for the case when $\phi_0=-0.4$ and $r=-0.9$, i.e., $\Delta=-0.42$. We did not extract the value of $k_{i}$ from the numerical results. However, as can be seen from the inset to (c), the agreement between the slope of the dashed line, which has gradient $-k_{i}$ as obtained from the full theory, and the envelope of the order parameter profile, is good. Panel (d) shows the corresponding results for the wave number $k^*$ of the nonlinear state deposited by the front (symbols) for comparison with the predicted wave number $k^*$ (solid line). The inset shows a plot of both $k^*$ and $k_r$ which confirms that the wave number behind the front differs from $k_r$, the wave number amplified by the front (dashed line). Both $k^*$ and $k_r$ differ substantially from $q=1$.}
\end{figure*}

We now assume $\Delta<0$ and calculate the speed with which the solidification front propagates into the unstable liquid. Taking the approximate dispersion relation in Eq.\ \eqref{eq:disp_rel_PFC} together with Eqs.\ \eqref{eq:speed1} and \eqref{eq:speed2}, hereafter the {\it full theory}, we obtain three equations for the three unknowns $c$, $k_r$ and $k_{i}$ \cite{GE11}. Two of the resulting equations are quintics in $k_r$ and $k_{i}$ and the third has a term in $k_r^6$. These simultaneous equations may be solved numerically. Results for the front speed $c$ obtained from doing this are displayed in Fig.\ \ref{fig:speed} (a) as a solid line. However, one can proceed further analytically by noting that when $\Delta$ is small $k_{i}$ is also small. We also make the ansatz that $k_{r}\approx q+ak_{i}$, where the constant $a$ is a variable to be solved for. We now proceed by expanding the three equations we obtain from Eqs.~\eqref{eq:speed1} and \eqref{eq:speed2} in powers of $k_{i}$. One can linearise all three equations in $k_{i}$ and then solve for $c$, $k_{i}$ and $a$ to obtain the following:
\begin{eqnarray}
c&=&\alpha q\sqrt{-8\Delta q^4+2\Delta^2}\\
k_{i}&=&\frac{q\sqrt{-8\Delta q^4+2\Delta^2}}{2(4q^4+\Delta)}\\
a&=&-\frac{2\Delta}{ \sqrt{-8\Delta q^4+2\Delta^2}}.
\end{eqnarray}
In addition, expanding Eq.~\eqref{eq:kbehind} yields the prediction
\begin{equation}
k^*=k_r-\frac{2\alpha}{c}qk_i(\Delta+4q^3ak_i-6q^2k_i^2)
\end{equation}
for the wave number behind the front.

These results show that when $\Delta$ is small, the front propagation speed $c\propto \sqrt{-\Delta}$. One also sees that $k_{i}\propto \sqrt{-\Delta}$, $a\propto\sqrt{-\Delta}$ while $k^*-k_r\propto |\Delta|$ and so increases as $-\Delta$ increases. The above results are accurate when $|\Delta|$ is small, but are not reliable when the system is deeply quenched, i.e., when $|\Delta|$ is not small. In particular, when this is the case, it is important to distinguish between the wave number $k_r$ predicted by the marginal stability condition and the wave number $k^*$ left behind by the moving front. In fact, one can obtain an expression for the crystallisation front speed $c$ that is more accurate for a larger range of values of $\Delta$ as follows. We start by linearising the real part of Eq.\ \eqref{eq:speed1} in $k_{i}$ to obtain
\begin{equation}
k_{i}=-\frac{q\Delta}{a(\Delta+4q^4)}.
\label{eq:k_im}
\end{equation}
We next expand the imaginary part of Eq.\ \eqref{eq:speed1} to second order in $k_{i}$ and use Eq.\ \eqref{eq:k_im} to obtain
\begin{equation}
a=-\frac{2\alpha\Delta q(16q^8-28q^4\Delta+\Delta^2)}{c(16q^8+8q^4\Delta+\Delta^2)}.
\label{eq:a}
\end{equation}
Together these results determine an approximation for $k_{r}\equiv q+ak_{i}$. Finally, we expand Eq.\ \eqref{eq:speed2} to second order in $k_{i}$. Using Eqs.\ \eqref{eq:k_im} and \eqref{eq:a} leads to the following expression for the crystallisation front speed:
\begin{equation}
c=\frac{4\alpha q^3(16q^8-28q^4\Delta+\Delta^2)\sqrt{-\Delta(\Delta^2-64q^4\Delta+16q^8)}}{(\Delta^2-64q^4\Delta+16q^8)(4q^4+\Delta)}.
\label{eq:analytic_speed}
\end{equation}

\subsection{Comparison with numerical simulations}
\label{sec:numsim}

In Fig.\ \ref{fig:speed}(a) we display the result from Eq.~\eqref{eq:analytic_speed} as the dashed line, together with the result from the full numerical solution (solid line) obtained from Eqs.\ \eqref{eq:speed1}, \eqref{eq:speed2} and \eqref{eq:disp_rel_PFC}. We see that for small values of $|\Delta|<0.1$ the expression in Eq.\ \eqref{eq:analytic_speed} for $c$ is accurate. However, for larger values of $|\Delta|$, it becomes less reliable. This approach also captures fairly well the behaviour of $k_{i}$, as can be seen in Fig.\ \ref{fig:speed}(c). However, as can be seen in (b), it does not describe very well the behaviour of $k_r$ as a function of $\Delta$.

In Fig.\ \ref{fig:speed}(a) and (b) we also display results for the front speed $c$ and the wave number $k_r$ obtained numerically by solving the PFC equations Eqs.\ \eqref{eq:DDFT_PFC} and \eqref{eq:freeenergy_PFC} on a 1-dimensional grid. We set the system size to be $\ge1000$, which is sufficient for a stationary advancing front to develop \footnote{To determine the front speed from 1D numerical simulations we used the following procedure: we typically calculate the profile on a system of length 2000 with periodic boundary conditions. The initial order parameter profile is uniform with value $\phi_0$ except for a single peak on the central grid point with $\phi=2\phi_0$ and the two grid points either side of the peak where $\phi=\phi_0/2$, to ensure that the average value in this region remains $\phi_0$. We then focus on one half of the domain, since the resulting structures are symmetric about the mid-point. We define the position of the interface as the point closest to the boundary where $|\phi-\phi_0|>10^{-5}$. We then run the simulations until at the end point of the domain $|\phi-\phi_0|>10^{-20}$. To calculate the front speed we first determine the time at which the interface reaches the point a distance 75 from where the front was initiated (to eliminate the effect of initial transients) and then locate the position of the interface at the end of the simulation and the time taken. From these measurements we obtain the front speed $c$. We calculate $k_r$ by calculating the distance between the peaks in $\phi$ in the traveling front region. To do this one must define a cut-off point, where the distance between peaks starts to crossover from the value $2\pi/k_r$ to the value $2\pi/k^*$. The distance between peaks in the front region is then defined as the distance between the first peak at the front of the interface and the last peak before this cut-off point, divided by the number of peaks between these two points.}. We compare results obtained for various values of the spatial grid spacing $dx$. We present results for $dx=0.2$, 0.5, 1 and $\pi/3$ (in the literature there are some groups that use this particular value). We find that for the larger values of the lattice spacing $dx$ the front speed $c$ is markedly slower than for smaller values of the lattice spacing, which are in good agreement with the exact speed obtained by solving Eqs.\ \eqref{eq:disp_rel_PFC}, \eqref{eq:speed1} and \eqref{eq:speed2} numerically and displayed as the solid line in Fig.\ \ref{fig:speed}(a). Finally, in Fig.\ \ref{fig:speed}(d) we display the corresponding results for $k^*$ and compare these with the theoretical predictions for $k^*$ (solid line) and $k_r$ (dashed line). The theoretical predictions for $c$, $k_r$ and $k^*$ (solid lines in Figs.~\ref{fig:speed}(a,b,d)) are in excellent agreement with the numerical results obtained with grid spacing $dx=0.2$. Figure \ref{fig:speed} also shows that results obtained with $dx>0.5$ are substantially in error. This is because the discretisation of the system effectively adds a friction term proportional to the magnitude of $dx$ to the dynamical equations, which slows down the advancing front -- i.e., the numerical grid can `pin' the advancing front. Evidently, this pinning effect is also reflected in the corresponding values of $k_r$ and $k^*$. 

We did not extract the value of $k_{i}$ from the numerical results. However, in the inset of Fig.\ \ref{fig:speed}(c) we display the order parameter profile plotted as $\log |\phi(x)-\phi_0|$ versus $x$ for the case when $\phi_0=-0.4$ and $r=-0.9$, calculated numerically using the grid spacing $dx=0.2$. From the analysis of the advancing front profile one expects that $\phi(x,t) -\phi_0 \sim \exp(-k_{i}(x-ct))\sin(k_r x)$, so that when the order parameter profile is plotted in this manner, the envelope function $\exp(-k_{i} x)$ of the advancing front profile becomes a straight line with gradient $-k_{i}$. The dashed line in the inset of Fig.\ \ref{fig:speed}(c) is a straight line with gradient $-k_{i}$ computed from Eqs.\ \eqref{eq:disp_rel_PFC}, \eqref{eq:speed1} and \eqref{eq:speed2}. It is clear that the gradient of the envelope of the numerically obtained order parameter profile is very close to that of the dashed line. Thus, we conclude that the analysis based on Eqs.\ \eqref{eq:disp_rel_PFC}, \eqref{eq:speed1} and \eqref{eq:speed2} leads to a prediction for the solidification front speed $c$ which is precisely that which one obtains from solving the PFC equations \eqref{eq:DDFT_PFC} and \eqref{eq:freeenergy_PFC}.

\begin{figure*}
\includegraphics[width=3in]{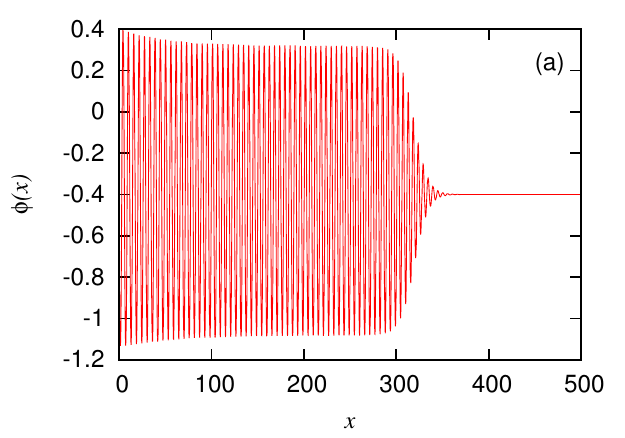} 
\includegraphics[width=3in]{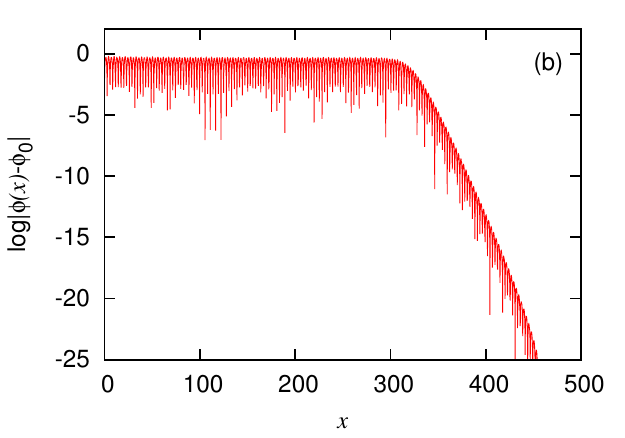} 
\includegraphics[width=3in]{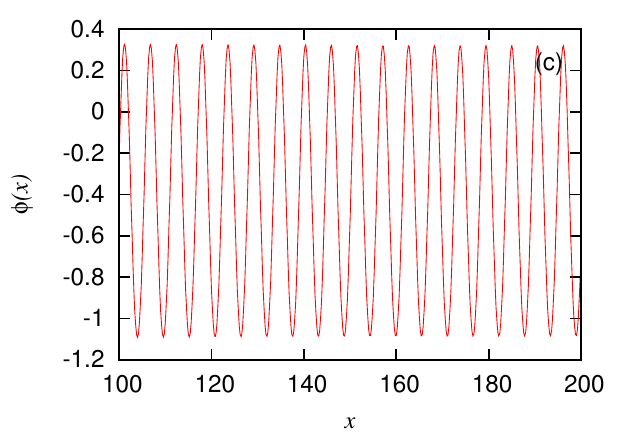}
\includegraphics[width=3in]{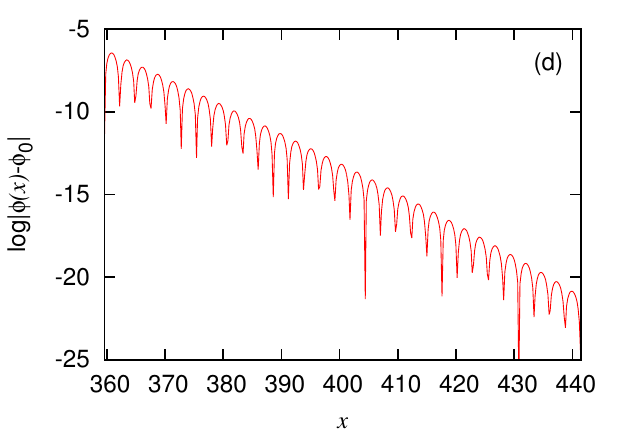}  
\caption{\label{fig:kbehind}
(Colour online) (a) A front advancing to the right at one instant of time when $r=-0.9$, $\phi_0=-0.4$, $q=1$ and $\alpha=1$, computed with $dx=0.2$. (b) The solution in panel (a) on a semi-logarithmic plot. (c) Enlargement of the region behind the front in panel (a). (d) Enlargement of the front region in panel (b).}
\end{figure*}

In Fig.~\ref{fig:kbehind} we show an example of a front propagating towards the right when $\Delta= -0.42$. The front region is clearly visible on the semi-logarithmic plot shown in panel (b); panels (c) and (d) show enlargements corresponding to the region behind the front and the front region itself. From these figures one determines that the wave number in the front region is $k_r\approx 1.189$, while the wave number behind the front is $k^*\approx 1.123$. These measurements agree very well with the exact marginal stability result, $k_r\approx 1.187$, and the prediction in Eq.~\eqref{eq:kbehind}, $k^*\approx 1.129$.

It is of interest to note that the dynamically selected wave number $k^*$, which determines the wavelength $\lambda=2\pi/k^*$ of the density modulations left behind the advancing front, can differ significantly from the equilibrium wavelength $\lambda_c\approx 2 \pi/q$ of the fully formed crystal. This means that for large negative values of $\Delta$, which corresponds to a deep quench (i.e., the unstable liquid is strongly supercooled), the system must perform significant rearrangements after the initial solidification front has passed, in order to obtain density modulations with wavelength $\approx 2 \pi/q$, corresponding to an ordered crystal of minimal energy. However, one should expect that these later rearrangements (ageing) are frustrated by the fact that the system has already chosen a different and dynamically selected length scale. As the system ages some of these defects anneal, reducing the disorder in the solid and bringing it closer to equilibrium. { We thus believe that the difference between the dynamic and equilibrium crystalline wavelengths may be an important factor in understanding why some rapidly quenched liquids and soft matter systems exhibit disorder rather than forming a regular crystalline material}. We illustrate and demonstrate this result further in the next section.

\section{PFC results}
\label{sec:PFC_res}

\begin{figure*}
\includegraphics[width=2.\columnwidth]{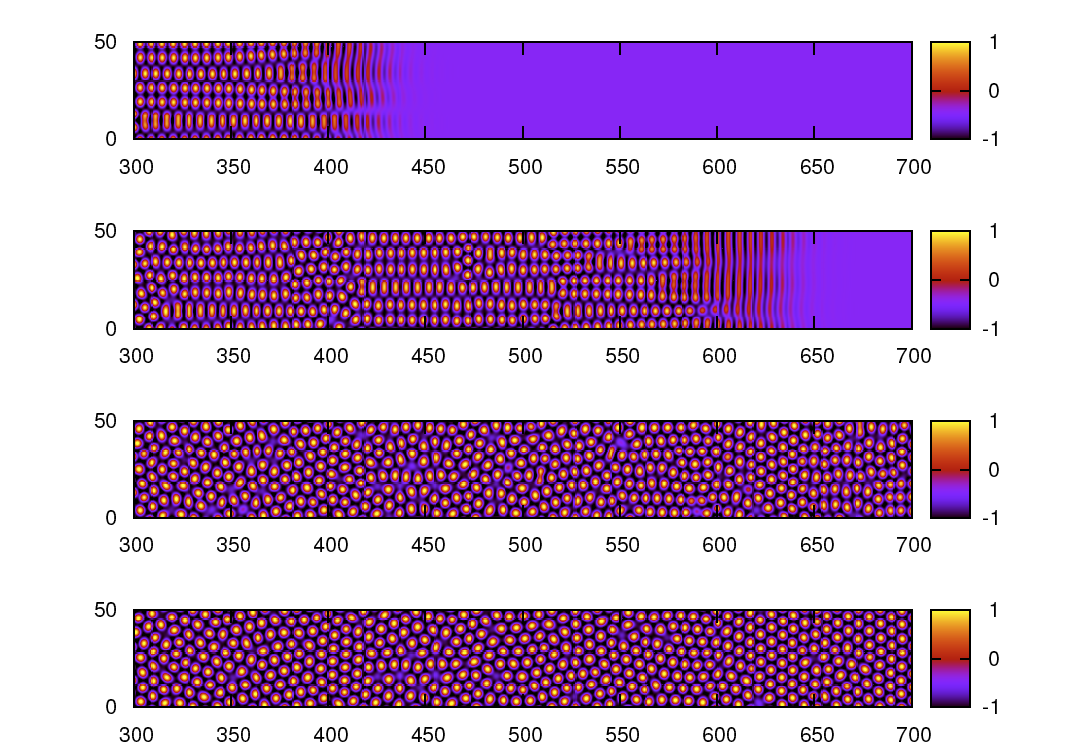}

\caption{\label{fig:profiles}
(Colour online) Order parameter profiles for a crystallisation (solidification) front advancing into the unstable uniform phase, obtained from the PFC model (Eqs.\ \eqref{eq:DDFT_PFC} and \eqref{eq:freeenergy_PFC}) in two spatial dimensions when $\phi_0=-0.43$ and $r=-0.9$, corresponding to $\Delta=-0.35$. The plots correspond to the times $t^*\equiv \alpha t/q^2=140$, 200, 260 and 340, going from top to bottom. The solidification front was initiated at $t=0$ at $x=0$ and propagates towards the right. Note the rearrangements that occur at points well behind the moving front.}
\end{figure*}

\begin{figure*}
\includegraphics[width=0.99\columnwidth]{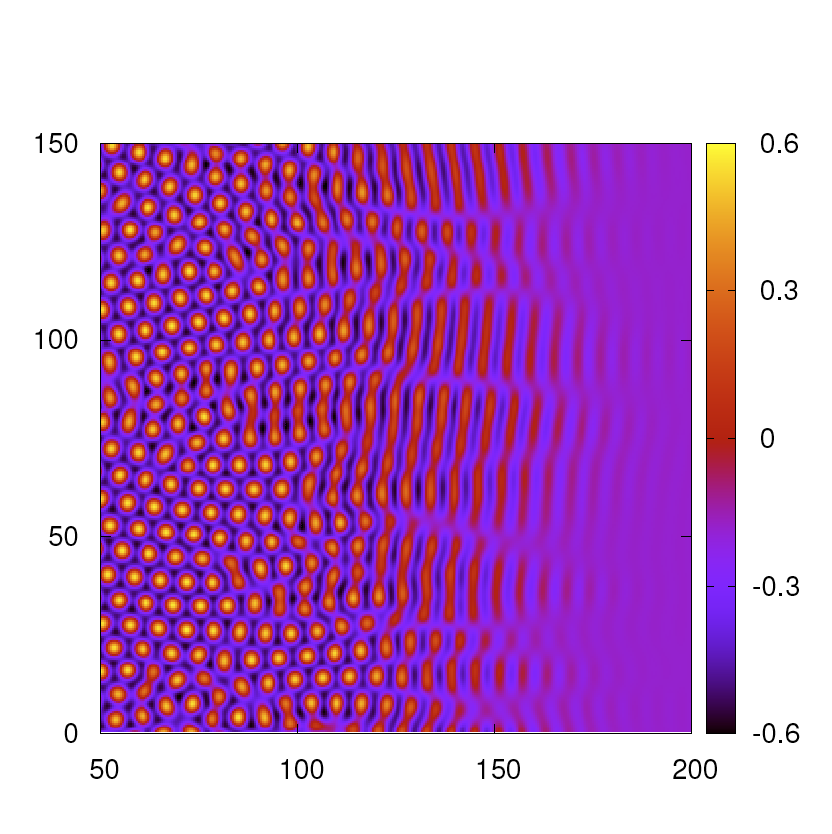}
\includegraphics[width=0.99\columnwidth]{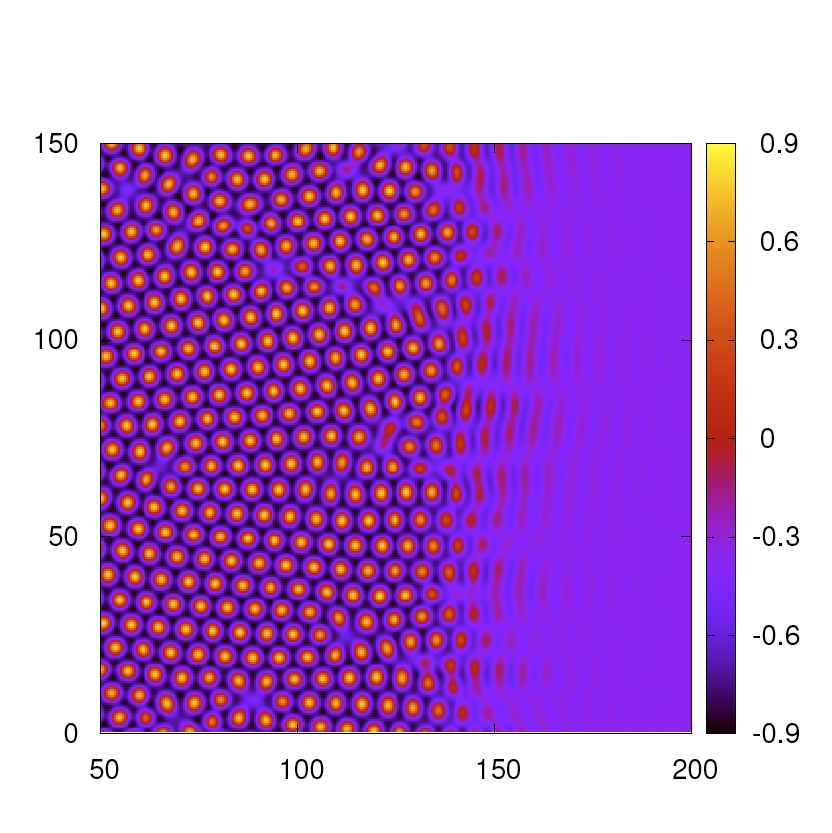}

\includegraphics[width=0.99\columnwidth]{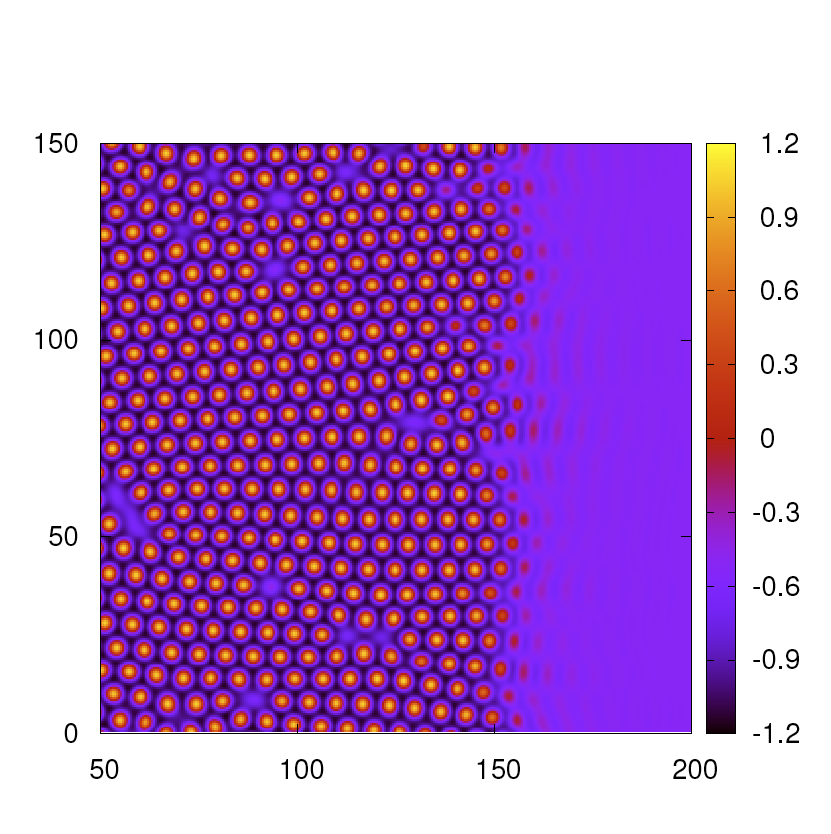}
\includegraphics[width=0.99\columnwidth]{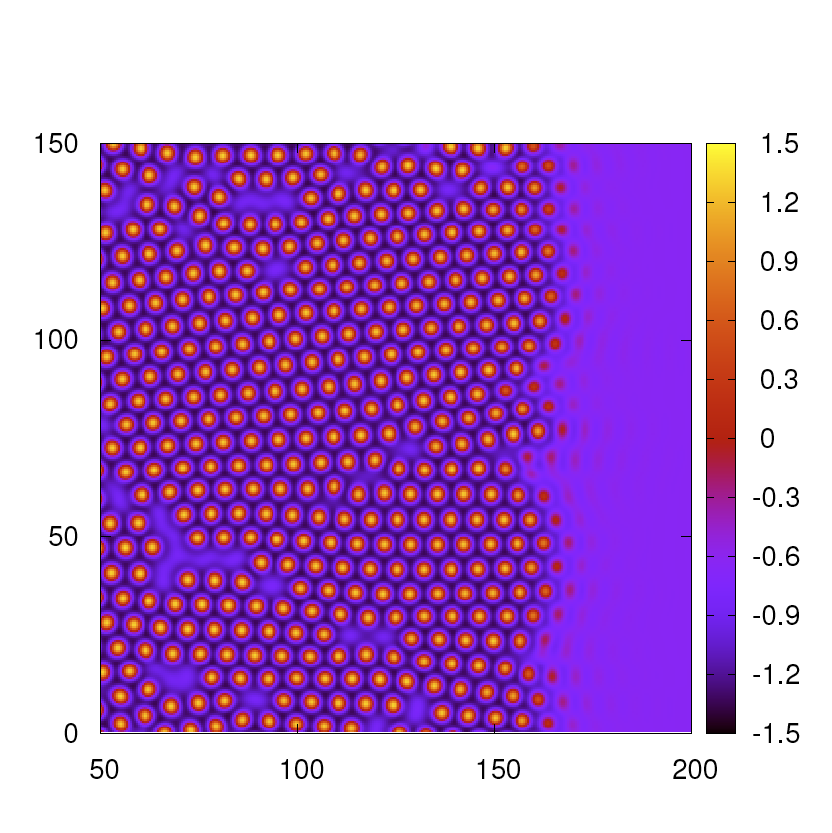}

\caption{\label{fig:profiles_fig7}
(Colour online) Order parameter profiles of a crystallisation (solidification) front advancing into the unstable uniform phase, obtained from the PFC model (Eqs.\ \eqref{eq:DDFT_PFC} and \eqref{eq:freeenergy_PFC}) in two spatial dimensions for $\Delta=-0.1$ and several different values of $r$, all at time $t^*\equiv \alpha t/q^2=152$. The solidification front was initiated at $x=0$ at $t=0$ and propagates towards the right. The top left panel is for $r=-0.2$ and $\phi_0=-0.183$, top right for $r=-0.5$ and $\phi_0=-0.365$, bottom left for $r=-0.9$ and $\phi_0=-0.516$ and bottom right for $r=-1.3$ and $\phi_0=-0.632$. The displayed region is part of a larger system of size $400\times400$ and calculated with a grid spacing $dx=dy=0.5$. Note the different colour table scales in each panel.}
\end{figure*}

In this section we confirm and illustrate the results and conclusions of the analysis given in the previous section, using results obtained from direct numerical simulations in two spatial dimensions for the simple PFC model given in Eqs.\ \eqref{eq:DDFT_PFC} and \eqref{eq:freeenergy_PFC}. This model system is now well understood and much is known about its thermodynamics and phase behaviour, and the structures that are formed \cite{EG04, BGE06, EPBSG07, BRV07, BEG08a, BEG08, MKP08, TBVL09, GE11, TTG11, TGTDP11}. We display in Fig.\ \ref{fig:profiles} the order parameter profiles for a solidification front advancing from left to right into the unstable uniform liquid phase, for a system with $q=1$, $\phi_0=-0.43$ and $r=-0.9$, corresponding to $\Delta=-0.35$. The profiles in Fig.\ \ref{fig:profiles} are calculated by taking an initially uniform system with $\phi(\xx)=\phi_0$, of size $2000\times50$ with periodic boundary conditions and grid spacing $dx=dy=0.5$. The solidification is initiated by adding small amplitude random noise to the profile along the line $x=0$ at time $t=0$. The order parameter profiles displayed in Fig.\ \ref{fig:profiles} correspond to the times $t^*\equiv \alpha t/q^2=140$, 200, 260 and 340. We see that the front advances by first forming stripe-like density modulations in the direction of travel, as predicted by the analysis in Sec.\ \ref{sec:front_speed} above. However, the stripes are typically broken into transverse domains or `filaments' (see Sec.~\ref{transverse}), leading to a two-dimensional structure that subsequently breaks up into density peaks resembling a solid. Figure~\ref{fig:profiles} shows the order parameter profile corresponding to the time $t^*=260$, and reveals that there is a significant amount of disorder in the arrangements of the density peaks shortly after the solidification front has passed. Then, over time, the system rearranges (ageing) leading to the more regular ordering seen in the order parameter profile for the time $t^*=340$ (see Sec.~\ref{ageing}).

The results of the PFC model depend on both the chosen value of $\Delta<0$, the undercooling, and of $\phi_0$, the background homogeneous state into which the solidification front propagates. To explore the parameter space, we solved the PFC model of size $400\times400$ with periodic boundary conditions in the $y$ direction and $dx=dy=0.5$, initializing the solidification front by adding small amplitude random noise to the order parameter profile along the line $x=0$ at the time $t=0$ \footnote{{ The small width $50\times2000$ system (Fig.\ \ref{fig:profiles}) is only used for illustrative purposes since it exhibits finite size effects; all other results are obtained using larger $400\times 400$ systems that have no discernible finite size effects.}}. We use the same realization of the initial condition throughout. According to the theory presented in the previous section, the front speed $c$ and wavenumbers $k_r$ and $k_i$ are determined by the value of $\Delta$ only. Figure \ref{fig:profiles_fig7} shows the results for $\Delta=-0.1$ and several different values of $r$ (equivalently of $\phi_0$, since $\phi_0=\sqrt{(\Delta-r)/3}$), all at the same time $t^*=152$ after the front was initiated at $x=0$ at $t=0$. Detailed analysis shows that the front speed and length scale right in the front region are indeed independent of the background value of $\phi_0$. On the other hand, the extent of the region of the stripe-like state is dramatically reduced as $|\phi_0|$ increases. Since mature stripes of wavelength $2\pi/k^*$ are created at the rate $\Im(\Omega)$ and destroyed at the rate $\omega_{\rm hex}$ at which the instability to hexagons manifests itself, it follows that the width $\ell$ of the stripe region scales as $\ell\sim 2\pi\Im(\Omega)/k^*\omega_{\rm hex}=2\pi c/\omega_{\rm hex}$ using Eq.~(\ref{eq:kbehind}), cf.~\cite{csahok99,hari00}. As shown in Appendix B, this quantity scales like $|\phi_0|^{-1}$ with a coefficient of proportionality that is independent of $\Delta$ when $|\Delta|\ll1$. Our numerical results are consistent with this prediction, although it is somewhat difficult to determine precisely the width $\ell$ from the data. In fact, simulations starting from random initial conditions show that for low $|\phi_0|$ the instability of the stripe state generates structures that are more rhomboid than hexagonal. With increasing $|\phi_0|$ the structures become more hexagonal but the fraction of vacancies within the structure goes up. This is a consequence of the fact that the curve $\Delta(\phi_0)=-0.1$ in the $(\phi_0,r)$ plane moves as $\phi_0$ increases and eventually crosses into the coexistence region between the hexagonal crystal and the homogeneous or liquid state \cite{RATK12}.

\subsection{The transverse length scale}\label{transverse}

\begin{figure}
\includegraphics[width=1.\columnwidth]{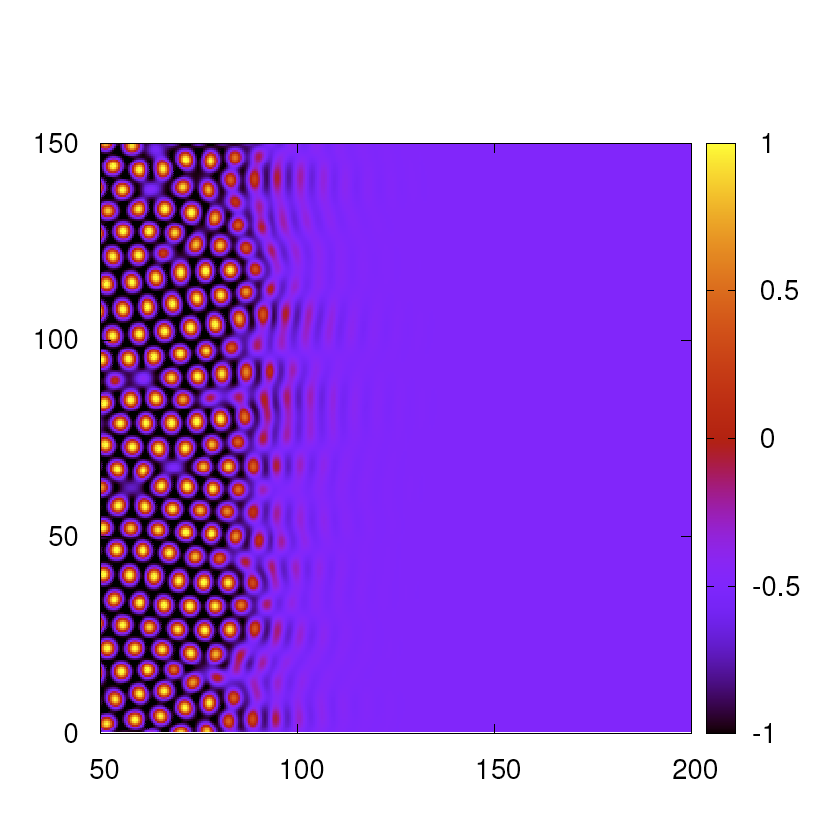}

\includegraphics[width=1.\columnwidth]{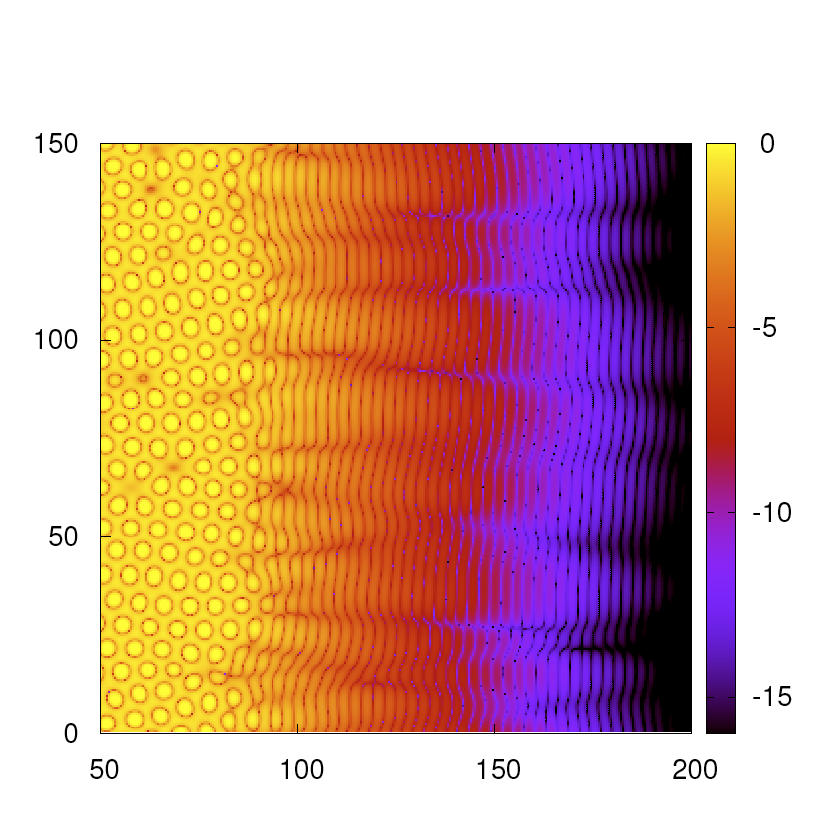}

\caption{\label{fig:profiles_log_comp}
(Colour online) Order parameter profile at time $t^*\equiv \alpha t/q^2=64$ for a crystallisation (solidification) front advancing into the unstable uniform phase, obtained from the PFC model (Eqs.\ \eqref{eq:DDFT_PFC} and \eqref{eq:freeenergy_PFC}) in two spatial dimensions when $\phi_0=-0.483$ and $r=-0.9$, corresponding to $\Delta=-0.2$. The solidification front was initiated at $x=0$ at $t=0$ and propagates towards the right. The upper figure shows the order parameter profile $\phi({\bf x})$, while the lower figure shows the same profile, but instead plotting the quantity $\ln|\phi({\bf x})-\phi_0|$. Plotting this quantity reveals the fine structure in the profile ahead of the front -- note the scale: the smallest amplitude structures that are displayed have an amplitude $\approx e^{-16}$). The displayed region is part of a larger system of size $400\times400$ and calculated with a grid spacing $dx=dy=0.5$.}
\end{figure}

Figures \ref{fig:profiles} and \ref{fig:profiles_fig7} reveal the presence of unambiguous filamentation of the stripe pattern created by the passage of the front. To understand the origin of this filamentation we show in Fig.~\ref{fig:profiles_log_comp}(a) the quantity $\phi(\xx)$ at time $t^*\equiv \alpha t/q^2=64$ when $\Delta=-0.2$ ($\phi_0=-0.483$ and $r=-0.9$) while Fig.~\ref{fig:profiles_log_comp}(b) shows the same solution but in terms of the quantity $\ln |\phi(\xx)-\phi_0|$. The latter representation not only rectifies the solution, but also amplifies it strongly in regions where $\phi(\xx)\approx\phi_0$. The figure reveals that the filamentation is present already at the front where the amplitude of the stripes is still minute, of order $e^{-16}$. Careful study of the origin of this filamentation shows that it is a consequence of the perturbation used to initialize the simulation. The ridges that break up the stripe pattern correspond to zero-crossings in $\phi(x=0,y,t=0)-\phi_0$, here a particular realization of a uniformly distributed random variable on the interval $[\phi_0-0.1,\phi_0+0.1]$. The regions where $\phi(x=0,y,t=0)-\phi_0\approx0$ travel more slowly than regions where $\phi(x=0,y,t=0)-\phi_0\ne0$, and the latter are broad enough to trigger the formation of stripe segments. Thus the filaments are an imprint of the initial condition, and the advancing front acts as a noise amplifier. Simulations initialized from a small amplitude perturbation with a single wavenumber $k_\perp$ preserve this wavenumber into the nonlinear regime and the resulting filamentation is periodic with wavelength $\lambda_\perp \equiv 2\pi/k_\perp$.

The stripes created by the advancing front are unstable to oblique disturbances that favour the formation of hexagonal structures and this instability becomes visible once $\phi-\phi_0=O(\phi_0)$ (Fig.~\ref{fig:profiles_log_comp}(a,b)). The growth rate of this instability is proportional to $|\phi_0|$ (see Appendix B) and consequently we expect the width $\ell$ of the stripe interval ahead of the hexagonal pattern to decrease with increasing $|\phi_0|$, all other parameters remaining fixed (cf.~Fig.~\ref{fig:profiles_fig7}). Our simulations reveal, however, that the filamentation imprinted by the initial conditions also has a strong effect on the ability of the system to form hexagons. If the characteristic transverse scale $\lambda_\perp$ is far from $2\lambda_\parallel/\sqrt{3}$, where $\lambda_\parallel\equiv 2\pi/k^*$, we find that the formation of hexagons is delayed until such time as the required wavenumber is generated by nonlinear interactions. Thus the initial condition strongly influences, through the above process, the time required to form the crystalline state. Moreover, since the selected wavelength $\lambda_\parallel$ is likewise non-optimal, both factors contribute to frustration and disorder in the solidification process for deep quenches. 

\subsection{Structure and correlations over time: ageing}\label{ageing}

\begin{figure*}
\includegraphics[width=0.67\columnwidth]{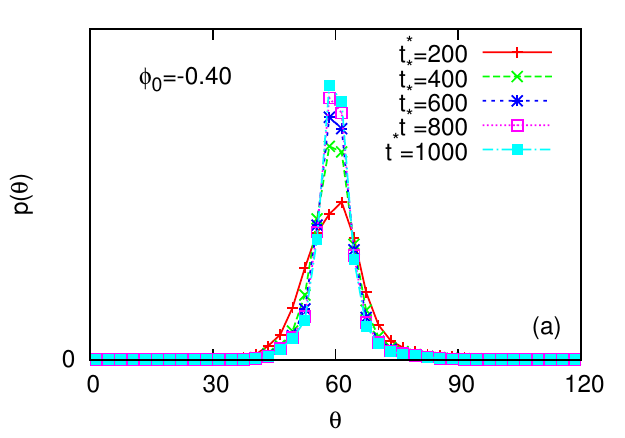}
\includegraphics[width=0.67\columnwidth]{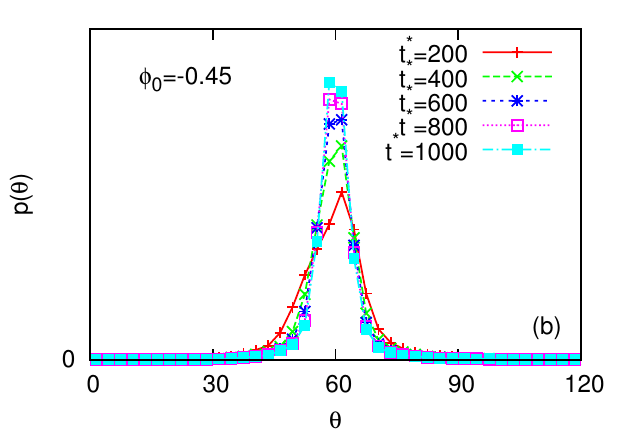}
\includegraphics[width=0.67\columnwidth]{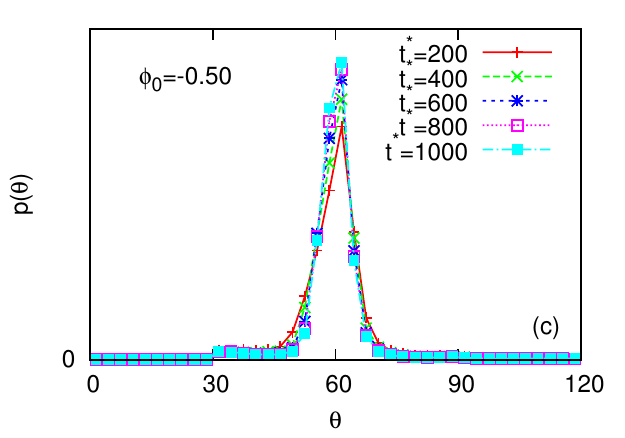}

\includegraphics[width=0.67\columnwidth]{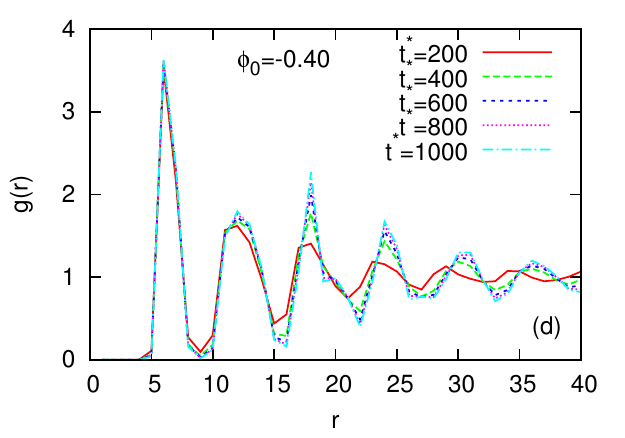}
\includegraphics[width=0.67\columnwidth]{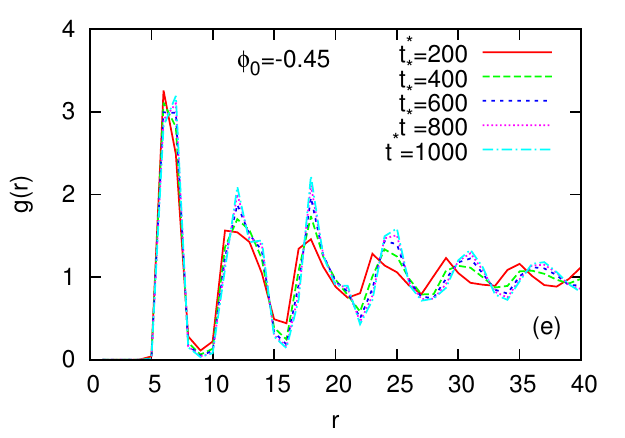}
\includegraphics[width=0.67\columnwidth]{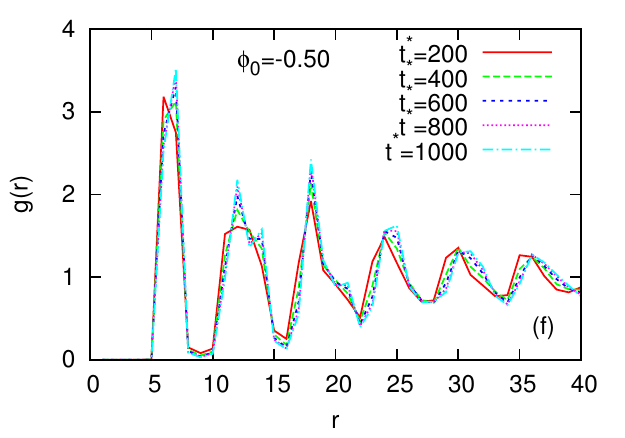}

\caption{\label{fig:Angle_and_g_of_r} (Colour online) The bond angle distribution $p(\theta)$ (top row) and radial distribution function $g(r)$ (bottom row) at various times $t^*$, after the solidification front was initiated. The undercooling parameter is $r=-0.9$, and the value of $\phi_0$ decreases from left to right (as indicated in the figures) resulting in (from left to right) $\Delta=-0.42$, $-0.29$ and $-0.15$. After the initial crystallisation front passes by, the system undergoes `ageing' as the particles are able to perform some rearrangements.}
\end{figure*}

In order to quantify the degree of order in the system and compare results from shortly after the solidification front has passed with those at a later time, we computed the bond angle distribution $p(\theta)$ and radial distribution function $g(r)$ as a function of time after the solidification front was initiated. These quantities are calculated from { larger} scale (grid size $400\times400$) simulations by first locating all the maxima in the order parameter profile, i.e., the coordinates of all the density peaks (particles) after the crystallisation front has moved through the system. From these sets of particle coordinates, we calculate the radial distribution function $g(r)$ in the usual way \cite{AllenTildesley}. Since $g(r)$ is a spatial two-point correlation function, it gives the probability of finding another particle at a distance $r$ away from any other given particle \cite{HM}. The bond angle distribution function is calculated by performing a Delauney triangulation on the system. The histogram of the values of the corner angles of this set of triangles (i.e., the nearest neighbour bond angles) is $p(\theta)$.

In Figs.\ \ref{fig:Angle_and_g_of_r}(a)--(c) we display the bond angle distribution $p(\theta)$ for $r=-0.9$ as it varies over time, for (a) $\phi_0=-0.4$, (b) $\phi_0=-0.45$ and (c) $\phi_0=-0.5$. These three values of $\phi_0$ correspond, respectively, to $\Delta=-0.42$, $-0.29$ and $-0.15$. These bond angle distributions are centred on the value $60^\circ$, due to the dominant hexagonal ordering in the system, and we see no peaks at $45^\circ$ and $90^\circ$, which would indicate square ordering \cite{RATK12}. We see in (a) and (b), corresponding to larger values of $|\Delta|$ (i.e., the deeper quenches), that at the time $t^*=200$ the distribution $p(\theta)$ is much broader than for later times, indicating that at this early time, shortly after the solidification front has passed through the system, there is much more disorder in the system than at the later times. Over time, the system rearranges to form a much more ordered solid, with $p(\theta)$ being much more sharply distributed around $60^\circ$. In contrast, for the shallow quench case with small $|\Delta|$ displayed in Fig.\ \ref{fig:Angle_and_g_of_r}(c), we see that $p(\theta)$ is sharply distributed around $60^\circ$ even for short times after the solidification front has moved through the system and that it does not change much as time goes by, indicating there is very little ageing in the system. These findings can also be seen by inspecting the radial distribution functions $g(r)$ displayed in Figs.\ \ref{fig:Angle_and_g_of_r}(d)--(f). For the shallow quench case in (f) we see that $g(r)$ does not change much over time. In contrast, for the deeper quench cases in (d) and (e) we see that at $t^*=200$ the decay $g(r) \to 1$ is much faster than at later times. The fact that the amplitude of the oscillations in $g(r)$ is much smaller at earlier times indicates that there is much less long range (crystalline) ordering in the system. As time proceeds, the amplitude of the oscillations in the tail of $g(r)$ grows, indicating that the system is rearranging to form a much more ordered system with the particle locations being well correlated over larger distances. The larger amount of disorder shortly after a deep quench is a consequence of the mismatch between the wavelength selected dynamically by the advancing solidification front and the equilibrium lattice spacing of the crystalline solid. This mismatch increases with increasing $|\Delta|$. The initial appearance of density modulation with the `wrong' wavelength creates disorder and frustration in the system, a picture corroborated by the results in Sec.\ \ref{sec:front_speed}.

\section{Concluding remarks}
\label{sec:conc}

In this paper we have studied the propagation of a solidification front into a supercooled liquid, i.e., into a linearly unstable state. We employed dynamical density functional theory (DDFT) to derive an approximate dispersion relation for small perturbations of the spatially uniform liquid state and noted that this dispersion relation is identical in form to that derived from the phase field crystal (PFC) model of crystal growth. In both approaches the solid phase is represented as a spatially structured state with local maxima in the density profile $\rho(\xx)$ or equivalently the order parameter $\phi(\xx)$ representing the time-averaged location of individual atoms/particles. The present approach is thus able to bridge purely continuum or macroscopic solidification theory \cite{umantsev85} with atomistic approaches such as molecular dynamics. Despite this fundamental difference, the DDFT and PFC models that result can still be formulated in terms of partial differential equations. These may be nonlocal as in DDFT or local as in PFC.

Knowledge of the dispersion relation suffices for the computation of the speed of the solidification front when this speed is selected by linear processes, i.e., in situations where the growth of the perturbations behind the front compensates for the propagation of the front, resulting in a steadily advancing front of constant shape. { However, in some problems the speed of the front may instead be determined by nonlinear processes \cite{saarloos03}.} For this reason it is essential to compare the prediction obtained from the linear marginal stability criterion employed here in the form of Eqs. \eqref{eq:speed1}--\eqref{eq:speed2} with numerical simulations. Such simulations yield in addition important information about processes occurring on longer time scales than the propagation time. Our results can be summarized as follows. For small undercooling, as measured by the parameter $|\Delta|$, the advancing front selects wavelengths close to the equilibrium wavelength $\lambda_c$ of the crystalline solid, resulting in steady transformation of the liquid state into solid. The front speed is $c\sim \sqrt{-\Delta}$. For large undercooling (i.e., supercooling) the front speed is faster and follows the approximate relation $c\sim -\Delta$. In this regime the wavelength selected by the advancing front differs substantially from $\lambda_c$ resulting in a nonequilibrium structure that subsequently evolves on a longer time scale, first via an instability to a hexagonal structure and subsequently via slow defect migration and annihilation. This `ageing' process consists of rearrangements as the system seeks to anneal out the defects and differently orientated domain structures which frustrate the formation of a regular crystal with wavelength $\lambda_c$.

We have also found that the initial perturbation imprinted on the advancing front may have a significant effect on the manifestation of the instability of the stripe state with respect to hexagonal perturbations. Since this transverse scale will also differ from the optimal scale $2\pi/k^*$ its presence provides an additional source of frustration following the passage of the front. Although these results were obtained using the PFC model, analogous two-dimensional calculations based on a DDFT model yield very similar results (not shown).

{ An important issue that we must mention concerns the extent to which insights from the PFC model can be applied to solidification in real materials. The transition in the PFC model from uniform to modulated phase is weakly first order, stemming from a truncated gradient expansion approximation to obtain the PFC free energy functional in Eq.\ \eqref{eq:freeenergy_PFC} (see also Eq.\ \eqref{eqFreeEngExApp2} in Appendix A). The fact that for some values of $\phi_0$ the PFC model exhibits a stripe phase that is not seen in real atomic fluids is an indication that the truncated gradient expansion approximation has failed for these $\phi_0$ values \cite{RATK12}. Thus, great caution should be taken in relating our results to solidification and glass formation in quenched liquids. For understanding how fronts propagate into a linearly unstable fluid, the approach described above appears to be valid. However, owing to the very simple nature of the PFC model, we expect that its description of the structures formed {\em behind} the front may be less reliable. The presence of a weakly first order transition in the PFC model makes it somewhat unrealistic as a model for materials like liquid metals, but for soft matter (polymeric) systems we believe it is a good approximation. Much more work comparing the PFC to more sophisticated DDFT approaches, such as that presented in Ref.\ \cite{TBVL09}, is required in order to elucidate the extent to which the PFC can be used to model real materials.

We mention, finally,} that the DDFT presented above was derived for Brownian particles. Improvements in the theory required for application to atomistic fluids include the DDFT \cite{archer17,archer25}:
\begin{equation}
\frac{\partial^2 \rho(\xx,t)}{\partial t^2}+\nu \frac{\partial \rho(\xx,t)}{\partial t}=\frac{1}{m} \nabla \cdot \left[ \rho(\xx,t)\nabla \frac{\delta F[\rho]}{\delta \rho(\xx,t)} \right],
\label{eq:DDFT_2}
\end{equation}
where $m$ is the mass of the atoms and  $\nu$ is the collision frequency given by $\nu\approx k_BT/mD$. Here $D$ is the self-diffusion coefficient. The free energy $F$ is given by \eqref{eq:freeenergy}. Front propagation in one-dimensional models of this type is considered in \cite{GE11}. Extensions of the present work to this class of models in two or more dimensions, together with a comparison with direct numerical simulations, will be presented elsewhere.

\acknowledgments

We acknowledge support by the EU via the ITN MULTIFLOW (PITN-GA-2008-214919). This collaboration was initiated while EK was a Visiting Professor in the Department of Mathematical Sciences at Loughborough University, funded by MULTIFLOW. AJA and MJR also acknowledge support from RCUK and EPSRC, respectively. 

\appendix

\section{Derivation of the PFC model from DDFT}
\label{sec5DDFT}

The DDFT in Eqs.\ \eqref{eq:DDFT} and \eqref{eq:freeenergy} is a microscopic theory which describes the time evolution of the fluid one-body (number) density profile $\rho(\xx,t)$ for a fluid of Brownian particles. In this section we start from the DDFT to derive the PFC model in its commonly used form. In our derivation we closely follow the arguments laid out in Ref.~\cite{TBVL09}. The excess contribution to the free energy $F_\mathrm{ex}$ in Eq.~\eqref{eq:freeenergy} is usually an unknown quantity.  Here, we make an approximation for $F_\mathrm{ex}$, by making a Taylor series expansion in powers of ${\tilde\rho}(\vx) = \rho(\vx) - \rho_0$, where $\rho_0$ is a reference density, giving \cite{Evan92}:
\begin{eqnarray}
	F_\mathrm{ex}[\rho(\vx)] &=& F_\mathrm{ex}[\rho_0] + \int \hspace{-1mm} d \vx \hspace{1mm}
	{\tilde\rho}(\vx) \fnDir{F_\mathrm{ex}[\rho(\vx)]}{\rho(\vx)} \evalAt{\rho_0} \nonumber \\   
	&& + \frac{1}{2} \int \hspace{-2mm} \int \hspace{-1mm} d \vx d \vx' \hspace{1mm} {\tilde\rho}(\vx) {\tilde\rho}
	(\vx') \fnDir{^2 F_\mathrm{ex}[\rho(\vx)]}{\rho(\vx) \delta \rho(\vx')} \evalAt{\rho_0} \nonumber \\
	&&+ O({\tilde\rho}^3).	
	\label{eqFexTaylor}
\end{eqnarray}
The functional derivatives of the excess free energy which enter into Eq.~\eqref{eqFexTaylor} are related to the $n$-body direct correlation functions in the following way \cite{Evan92}:
\begin{equation}
	\fnDir{^nF_\mathrm{ex}[\rho(\vx)]}{\rho(\vx_1) \delta \rho(\vx_2) \cdots \delta \rho(\vx_n)} \evalAt{\rho_0} 
	= -k_BT c^{(n)}(\vx_1,\vx_2, \cdots, \vx_n).
	\label{eqFnDirRule}
\end{equation}
In particular, the first member of this series is the one-body direct correlation function, shown earlier in Eqs.~\eqref{eq:Smoluchowski_2} and \eqref{eq:c_1_expansion}.  Note that the one body direct correlation function evaluated in the bulk is equal to the excess chemical potential $-k_BT c^{(1)}(\vx) \big|_{\rho_0} = \mu_\mathrm{ex}$. The second member of the series in Eq.~\eqref{eqFnDirRule} is the direct pair correlation function:
\begin{equation}
	\fnDir{^2F_\mathrm{ex}}{\rho(\vx) \delta \rho(\vx')} = -k_BTc^{(2)}(\vx,\vx'). 
	\label{eqTwoBodDirCorr}
\end{equation}
Substituting these expressions for the functional derivatives into Eq.~\eqref{eqFexTaylor} and neglecting third and higher order terms we obtain:
\begin{eqnarray}
	F_\mathrm{ex}[\rho(\vx)] \approx F_\mathrm{ex}[\rho_0] + \mu_\mathrm{ex} \int \hspace{-1mm} d \vx 
	\hspace{1mm}{\tilde\rho}(\vx) \hspace{1.5cm} \nonumber \\
	- \frac{k_BT}{2} \int \hspace{-2mm} \int \hspace{-1mm} d \vx d \vx' 
	\hspace{1mm} {\tilde\rho}(\vx)c^{(2)}(\vx,\vx'){\tilde\rho}(\vx')
	\label{eqFreeEngExApp}
\end{eqnarray}
The second term in this equation corresponds simply to a shift in the chemical potential and so this approximation is commonly used without the second term explicitly written down \cite{LLAB07, TLL08, TBVL09} as originally done by Ramakrishnan and Yussouff \cite{RaYu79}. To derive the PFC free energy, we make a gradient expansion of the two body direct correlation function and truncate at the fourth order term, giving \cite{TBVL09, EPBS07}:
\begin{equation}
	c^{(2)}(\vx,\vx') \approx -\beta(\hat{C}_0 + \hat{C}_2 \nabla^2 + \hat{C}_4 \nabla^4)\delta(\vx - \vx'),
	\label{eqTwoBodCorrApp}
\end{equation}
where in principle all the coefficients $\hat{C}_i$ are functions of $\rho(\vx)$, although we assume here that the coefficients $\hat{C}_2$ and $\hat{C}_4$ are in fact constants.  Inserting approximation \eqref{eqTwoBodCorrApp} into Eq.~\eqref{eqFreeEngExApp} gives:
\begin{eqnarray}
	F_\mathrm{ex}[\rho(\vx)] \approx F_\mathrm{ex}[\rho_0] + \mu_\mathrm{ex} \int \hspace{-1mm} d \vx 
	\hspace{1mm}{\tilde\rho}(\vx) \hspace{1.5cm}\nonumber \\
	+ \frac{1}{2} \int \hspace{-1mm} d \vx \hspace{1mm} {\tilde\rho}(\vx) 
	(\hat{C}_0 + \hat{C}_2 \nabla^2 + \hat{C}_4 \nabla^4) {\tilde\rho}(\vx),
	\label{eqFreeEngExApp2}
\end{eqnarray}
which makes $F_\mathrm{ex}[\rho(\vx)]$ a local functional.  Using this expression for the excess free energy term we can now write the Helmholtz free energy for the system as:
\begin{equation}
	F[\rho(\vx)] = \int \hspace{-1mm} d \vx \hspace{1mm} \bigg[ f_0(\rho(\vx)) + \frac{1}{2} {\tilde\rho}
	(\hat{C}_2 \nabla^2 + \hat{C}_4 \nabla^4) {\tilde\rho} \bigg], 
	\label{eqTempBF} 
\end{equation}   
where
\begin{equation}
	f_0(\rho) = k_BT \rho (\ln(\rho) - 1) + f_\mathrm{ex}[\rho_0]  +  \mu_\mathrm{ex} {\tilde\rho} + \frac{1}{2} \hat{C_0}(\rho) {\tilde\rho}^2.
\end{equation}
Here the first term in $f_0(\rho(\vx))$ comes from the ideal gas contribution (see Eq.~\eqref{eq:freeenergy}) and we have assumed that the external potential $V_{ext}=0$. We also make a further approximation by making a Taylor expansion of the function $f_0(\rho)$ around the reference density $\rho_0$, giving:
\begin{eqnarray}
	f_0(\rho) &\approx& f_0(\rho_0) + f_0'(\rho_0) {\tilde\rho} + \frac{f_0''(\rho_0)}{2}{\tilde\rho}^2 \nonumber \\
	&&+ \frac{f_0^{(3)}(\rho_0)}{3!} {\tilde\rho}^3 + \frac{f_0^{(4)}(\rho_0)}{4!} {\tilde\rho}^4.
	\label{eqLFTaylor}
\end{eqnarray}
We choose the reference density $\rho_0$ so that the third derivative of the function $f_0(\rho)$ vanishes at $\rho = \rho_0$, i.e.,~$f_0^{(3)}(\rho_0) = 0$. This gives the following:
\begin{equation}
	f_0(\rho) \approx f_0(\rho_0) + f_0'(\rho_0) {\tilde\rho} + \frac{f_0''(\rho_0)}{2}{\tilde\rho}^2 + \frac{f_0^{(4)}(\rho_0)}{4!} {\tilde\rho}^4.
	\label{eqTempLF}
\end{equation}
We now introduce a change of variables.  We use the non-dimensional variable 
$\phi = {{\tilde\rho}}/{\rho_1}$, where $\rho_1$ is a constant density, so Eqs.~\eqref{eqTempBF} and \eqref{eqTempLF} become:
\begin{equation}
	F[\phi(\vx)] = \int \hspace{-1mm} d \vx \hspace{1mm} \bigg[ f_0(\phi(\vx)) + \frac{1}{2} \phi 
	(C_2 \nabla^2 + C_4 \nabla^4) \phi \bigg] ,
	\label{eqTempBF2} 
\end{equation} 
where $C_2 = \hat{C}_2/\rho_1^2$, $C_4 = \hat{C}_4/\rho_1^2$ and
\begin{equation}
	f_0(\phi) \approx a + b\phi + \frac{c\phi^2}{2} + \frac{d\phi^4}{4},
	\label{eqTempLF2} 
\end{equation}
where $a$, $b$, $c$ and $d$ are constants.  

We now consider the dynamics of the model.  We start with the DDFT equation \eqref{eq:DDFT}. In the limit where $\rho_1 \phi$ is small, the density preceding the gradient of the functional derivative becomes constant, i.e.,~$\rho = \rho_0 + \rho_1 \phi \approx \rho_0$ and Eq.~\eqref{eq:DDFT} reduces to the following equation:
\begin{equation}
	\paDir{\rho(\vx, t)}{t} = \Gamma\rho_0 \nabla^2 \fnDir{F[\rho(\vx,t)]}{\rho(\vx,t)}.
\end{equation}
This is often referred to as ``model B" dynamics in the classification of Hohenberg and Halperin \cite{HoHa77}. Equivalently, we have the following equation for the time evolution of the order parameter $\phi(\vx, t)$:
\begin{equation}
	\paDir{\phi(\vx, t)}{t} = \alpha \nabla^2 \fnDir{F[\phi(\vx,t)]}{\phi(\vx,t)},
	\label{eqPFCDyn}
\end{equation}
where $\alpha = \Gamma \rho_0/\rho_1^2$ is the mobility coefficient.  Since the constant and linear terms in Eq.~\eqref{eqTempLF2} are irrelevant for the dynamics, we may drop the terms $a + b\phi$ from the function $f_0(\phi)$ in Eq.~\eqref{eqTempLF2}.  The functional derivative of the free energy is then given by the expression:
\begin{eqnarray}
	\fnDir{F}{\phi} 
	 =  d \biggl(\frac{c}{d} \phi + \phi^3 + \frac{C_2}{d} \nabla^2 \phi + \frac{C_4}{d} \nabla^4 \phi\biggr).
	\label{eqPFCFD}
\end{eqnarray}
We may absorb the parameter $d$ into the mobility coefficient $\alpha$. Also, we may choose $\rho_1$ so that $C_4/d = 1$. Writing $C_2/d = 2 q^2$ and $c/d = r + q^4$, we arrive finally at the commonly used PFC free energy:
\begin{equation}
	F[\phi(\vx)] = \int \hspace{-1mm} d\vx \hspace{1mm} f(\phi(\vx)),
	\label{eqPFCBF}
\end{equation}
where
\begin{eqnarray}
	f(\phi) &=& \frac{r + q^4}{2} \phi^2 + \frac{\phi^4}{4} + \frac{1}{2}\phi (2q^2 \nabla^2 + \nabla^4) \phi, 
	\nonumber \\ &=& \frac{\phi}{2} \big[ r+ (q^2 + \nabla^2)^2\big] \phi + \frac{\phi^4}{4}. 
	\label{eqPFCLF}
\end{eqnarray}
Inserting these parameter values into the functional derivative of the free energy (Eq.~\eqref{eqPFCFD}), we obtain $\fnDir{F}{\phi} = (r + q^4) \phi + \phi^3 + 2q^2 \nabla^2 \phi + \nabla^4 \phi$.  The PFC model is then given by the conserved dynamics in Eq.~\eqref{eqPFCDyn}, where the free energy is given by Eqs.~\eqref{eqPFCBF} and \eqref{eqPFCLF}.

\section{Instability of the stripe state}

In this Appendix we determine the timescale of the instability of the stripe state. This instability leads to the formation of the hexagonal structures shown in Figs.~\ref{fig:profiles_fig7} and \ref{fig:profiles_log_comp}.

We write the PFC model in the form
\begin{eqnarray}
{\tilde\phi}_t=\alpha\nabla^2[\Delta{\tilde\phi}+(q^2+\nabla^2)^2{\tilde\phi}+3\phi_0{\tilde\phi}^2+{\tilde\phi}^3],
\end{eqnarray}
where $\Delta\equiv r+3\phi_0^2$ and
\begin{eqnarray}
{\tilde\phi}&\equiv&\phi-\phi_0\nonumber \\
&=&Ae^{ikx}+Be^{ik(-x+\sqrt{3}y)/2}+Ce^{ik(-x-\sqrt{3}y)/2}\nonumber\\
&\,&+{\rm c.c.}+{\rm h.o.t.},
\end{eqnarray}
{ where c.c.~denotes the complex conjugate of the preceding terms and h.o.t.~denotes higher order terms.} Here $A$ is the small but complex amplitude of the longitudinal mode while $B$ and $C$ are the corresponding amplitudes of two symmetry-related oblique modes. The state $(A,B,C)=(A,0,0)$ thus corresponds to the stripe state while $(A,B,C)=(A,A,A)$ corresponds to the hexagon state, with $A>0$ representing a hexagonal array of spots and $A<0$ representing a hexagonal array of holes or vacancies. 

Weakly nonlinear theory now leads to the following equations for the amplitudes $A,B,C$:
\begin{eqnarray}\label{amp}
A_t&=&-\alpha k^2[{\widetilde\Delta}A+6\phi_0{\bar B}{\bar C}+\dots],\\
B_t&=&-\alpha k^2[{\widetilde\Delta}B+6\phi_0{\bar C}{\bar A}+\dots],\\
C_t&=&-\alpha k^2[{\widetilde\Delta}C+6\phi_0{\bar A}{\bar B}+\dots],
\end{eqnarray}
where the overbar denotes complex conjugation and ${\widetilde\Delta}\equiv \Delta+(q^2-k^{2})^2$ represents the bifurcation parameter shifted in proportion to the departure of the wavenumber $k$ away from its optimal value $k=q$. By applying appropriate translations we may take $A,B,C$ to be real. We also take $B=C$ in order to focus on the instability of the stripe state with respect to hexagon-forming perturbations. The linear instability of an $(A_0,0,0)$ state with respect to such perturbations is then described by the equation
\begin{eqnarray}
B_t=-\alpha k^2[{\widetilde\Delta}+6\phi_0A_0]B,
\end{eqnarray}
implying that the growth rate $\omega_{\rm hex}$ of the hexagon instability is given by
\begin{eqnarray}
\omega_{\rm hex}=-\alpha k^2[{\widetilde\Delta}+6\phi_0A_0]. 
\end{eqnarray}
Here $A_0$ is the amplitude of the stripe state. Within Eq.~(\ref{amp}) this amplitude is not determined: the growing stripe state (${\widetilde\Delta}<0$) does not saturate. However, the saturation amplitude of the stripe phase can be computed by setting $B=C=0$ and extending the above approach to cubic order while imposing the requirement that $\langle{\tilde\phi}\rangle=0$, where $\langle\cdots\rangle$ denotes an average over the domain. We obtain $A_0^2=-4{\widetilde\Delta}(3-2\phi_0^2/q^4)^{-1}$. For $\Delta\ll1$ these results (with ${\widetilde\Delta}$ replaced by $\Delta$) apply to stripes with $k=k^*$ since $k^*\approx q$.

Since $\Delta<0$ for instability of the liquid phase, and likewise $\phi_0<0$, the growth rate $\omega_\mathrm{hex}$ is positive for all $A_0>0$ with $k^*\approx q$, implying that the stripe state is always unstable with respect to the formation of the hexagon state with $A=B=C>0$, i.e., a hexagonal array of spots. In the case $\phi^2_0>3q^4/2$ the bifurcation to stripes is subcritical and the hexagon instability then competes with an amplitude instability. However, near threshold $k^*\approx q$ and the growth rate of the latter is therefore $O(|\Delta|)$ while the growth rate of the hexagon instability is $O(|\phi_0|\sqrt{|\Delta|})$ and so is larger. In either case, the longitudinal width $\ell$ of the band of stripes ahead of the hexagonal state is predicted to scale, for small $|\Delta|$, as $\ell\sim (2\pi c/\omega_{\rm hex})+\gamma_1\sim\gamma_0|\phi_0|^{-1}+\gamma_1$, where $\gamma_0$ is independent of $|\Delta|$ but $\gamma_1\propto k_i^{-1}$ does depend on $|\Delta|$. For larger $|\Delta|$ the approximation in Eq.~(\ref{eq:analytic_speed}) is useful.

The saturated hexagon state can be included selfconsistently in the above theory only when $|\phi_0|\ll1$, i.e., when $A^2\sim |\phi_0|A \sim {\widetilde\Delta}$ \cite{pismen94,doelman03}. This is not the case in our simulations and we do not pursue this approach.

\end{document}